# Diffusive non-reciprocity and thermal diode

Ying Li[1,2,3,4*], Jiaxin Li[5,4], Minghong Qi[1,2,3], Cheng-Wei Qiu[4,†], and Hongsheng Chen[1,2,3]

[1] Interdisciplinary Center for Quantum Information, State Key Laboratory of Modern Optical Instrumentation, College of Information Science and Electronic Engineering, Zhejiang University, Hangzhou 310027, China

[2] ZJU-Hangzhou Global Science and Technology Innovation Center, Key Lab. of Advanced Micro/Nano Electronic Devices & Smart Systems of Zhejiang, Zhejiang University, Hangzhou 310027, China

[3] International Joint Innovation Center, ZJU-UIUC Institute, The Electromagnetics Academy at Zhejiang University, Zhejiang University, Haining 314400, China

[4] Department of Electrical and Computer Engineering, National University of Singapore, Singapore 117583, Singapore

[5] State Key Laboratory of Robotics and System, Harbin Institute of Technology, Harbin 150001, China

*eleying@zju.edu.cn
†chengwei.qiu@nus.edu.sg

Wave propagation and diffusion in linear materials preserve local reciprocity in terms of a symmetric Green's function. For wave propagations, the relation between the fields entering and leaving a system is more relevant than the detailed information about the fields inside it. In such cases, the global reciprocity of the scattering off a system through several ports is more important, which is defined as the symmetric transmission between the scattering channels. When a two-port system supports non-reciprocal (electromagnetic, acoustic) wave propagation, it is a (optical, phonon) diode directly following the definition. However, to date no concrete definition or discussion has been made on the global reciprocity of diffusive processes through a multiple-port system. It thus remains unclear what are the differences and relations between the three concepts, namely local non-reciprocity, global non-reciprocity, and diode effect in diffusion. Here, we provide theoretical analysis on the frequency-domain Green's function and define the global reciprocity of heat diffusion through a two-port system, which has a different setup from that of a thermal diode. We further prove the equivalence between a heat transfer system with broken steady-state global reciprocity and a thermal diode, assuming no temperature-dependent heat generation. The validities of some typical mechanisms in breaking the diffusive reciprocity and making a thermal diode have been discussed. Our results set a general background for future studies on symmetric and asymmetric diffusive processes.

## I. Introduction

In processes such as mass transport, thermal conduction, and direct current (DC) transport, the evolution of physical fields at macroscale can be described with a diffusion equation. Recently, artificial functional materials have brought exciting new possibilities to control diffusive processes. For example, various metamaterials enable the cloaking of heat diffusion [1]-[7], DC current [8]-[13], and light diffusion [14]-[17]. In particular, thermal metamaterials [18] are also designed for many other functionalities including heat collection [19]-[23], temperature management [24], heat signal camouflage [25]-[28], and asymmetric heat transfer [29]-[30].

The device that transfers heat asymmetrically in opposite directions is called a thermal diode [31],[32],[33] or thermal rectifier [34], which plays an important role in temperature management and thermal information processing [35]. To build a thermal diode, simply combining linear materials is not enough, because heat diffusion is always symmetric in them. Such a constraint is believed to due to the microscopic reversibility, or the time-reversal symmetry of heat carrier dynamics, based on which the Onsager-Casimir relation [36],[37] requires that the thermal conductivity should be a symmetric tensor. Furthermore, the Green's function for the equation governing heat diffusion in linear materials must also be symmetric [38]. Namely, if one swaps the positions of a point heat source and the target point, the corresponding Green's function remains the same. We refer to this property as the local reciprocity of heat diffusion as the reference point can be any local point inside the material.

The time-reversal symmetry and Onsager-Casimir relation also apply on electromagnetic (EM) fields [39],[40], governing the symmetry of the permittivity, permeability, and bi-anisotropic coupling tensors. Similar as in heat diffusion, the Green's function for EM wave propagation in linear materials is symmetric, known as the Lorentz reciprocity. What is unique for EM or other wave propagation is that this local reciprocity in linear materials implies a global reciprocity of their combinations. That is, the scattering matrix of a linear system is symmetric. The scattering parameters directly relate the output and input signals, so the global reciprocity is much more important than local reciprocity in practical applications. Indeed, the recent active researches on non-reciprocal wave propagation are all about breaking the global reciprocity of certain devices [41],[42].

As mentioned above, thermal devices like thermal diode have equal importance as those in wave physics. However, there is still no clear definition of global reciprocity for thermal devices to distinguish it from the local version, not to say a criterion for its breaking. This is mainly due to the lack of related concepts including input and output (or incoming and outgoing) fields to construct a scattering theory for heat diffusion in analogue of wave propagation. Another problem is that the relation between global non-reciprocity and thermal diode is unclear. In wave physics, the asymmetric scattering parameters between two ports directly imply that signal should be transferred differently between them in opposite directions, which meets the property of an optical [43] or phonon [44] diode. In heat diffusion, it is not so straightforward even if the scattering parameters can be properly defined. That is because a thermal diode usually works with different temperatures maintained at both ends, instead of simply heating or cooling one end. As a result, whether a two-port thermal device with broken global reciprocity is always a thermal diode remains a question.

Here, we provide an in-depth study on the local and global reciprocities in the context of heat diffusion. By considering time-harmonic situations, the scattering theory for heat diffusion through ports is established. The corresponding conditions for the preserving and breaking of global reciprocity are formulated, with meaningful results in the limit of zero frequency when the fields cannot be decomposed to input and output parts. We further prove that any device that preserves global reciprocity cannot be a thermal diode, while that with global non-reciprocity is a thermal diode. Based on our findings, methods to break diffusive reciprocity and realize a thermal diode are summarized, with some typical examples illustrated.

## II. Local reciprocity of heat diffusion

In macroscopic heat transfer, the temperature field $T(\bm{r},t)$ ($\bm{r}$ is the position vector, and $t$ is time) follows

$$\rho c_p \frac{\partial T}{\partial t} = \nabla \cdot (\bm{\kappa} \cdot \nabla T) + \rho c_p \bm{v} \nabla T + h(\bm{r},t) \tag{1}$$

where $\rho$ is the density, $c_p$ is the specific heat capacity at constant pressure, $\kappa$ is the thermal conductivity tensor of the material, and $h(r,t)$ is the density of heat source in bulk. We also included a convective term for generality, where $v$ is the velocity. We assume forced convection and temperature-independent velocity field $v$ to avoid nonlinearity. In many situations, the steady state or steady oscillatory state is of interest, and it is convenient to study the Fourier transformation of Eq. (1)

$$i\omega\rho c_p T(r,\omega) = \nabla \cdot [\kappa \cdot \nabla T(r,\omega)] + \rho c_p v \cdot \nabla T(r,\omega) + h(r,\omega) \tag{2}$$

It is well known that when 1) the material is temperature independent; 2) the thermal conductivity tensor is symmetric; 3) there is no convection; and 4) the material properties do not vary with time, Eq. (2) preserves reciprocity in the sense that the Green's function is symmetric [38] $G(r_1|r_0,\omega) = G(r_0|r_1,\omega)$.

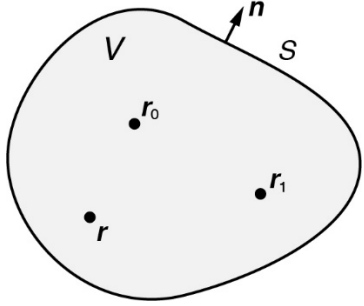

FIG. 1. Local reciprocity in heat transfer

Consider a domain $V$ as in Fig. 1, where all four conditions are satisfied. The Green's function follows

$$i\omega\rho c_p G(r | r_0,\omega) = \nabla \cdot [\kappa \cdot \nabla G(r | r_0,\omega)] + Q_0 \delta(r - r_0) \tag{3}$$

where $\delta(r)$ is the Dirac delta function, $Q_0$ is a constant to ensure correct unit. On the domain boundary $S$ with normal vector $n$, the Green's function satisfies the Dirichlet ($G(r_s|r_0,\omega) = 0$), Neumann ($[\kappa \nabla G(r_s|r_0,\omega)] \cdot n = 0$), or mixed ($cG(r_s|r_0,\omega) + [\kappa \nabla G(r_s|r_0,\omega)] \cdot n = 0$, where $c$ is a constant) boundary condition. Replacing $r_0$ with $r_1$ gives

$$i\omega\rho c_p G(r | r_1,\omega) = \nabla \cdot [\kappa \cdot \nabla G(r | r_1,\omega)] + Q_0 \delta(r - r_1) \tag{4}$$

By multiplying Eq. (3) with $G(r|r_1,\omega)$ and Eq. (4) with $G(r|r_0,\omega)$, then integrating their difference over $V$, we have

$$\int_V \{G(r | r_1,\omega)\nabla \cdot [\kappa \cdot \nabla G(r | r_0,\omega)] - G(r | r_0,\omega)\nabla \cdot [\kappa \cdot \nabla G(r | r_1,\omega)]\} dV$$
$$= \int_V [G(r | r_0,\omega)Q_0\delta(r - r_1) - G(r | r_1,\omega)Q_0\delta(r - r_0)] dV \tag{5}$$

The Gauss's law can be applied to the left-hand side

$$\text{l.h.s} = \int_V \nabla \cdot [G(r | r_1,\omega)\kappa \cdot \nabla G(r | r_0,\omega) - G(r | r_0,\omega)\kappa \cdot \nabla G(r | r_1,\omega)] dV$$
$$= \int_{\partial V} [G(r_s | r_1,\omega)\kappa \cdot \nabla G(r_s | r_0,\omega) - G(r_s | r_0,\omega)\kappa \cdot \nabla G(r_s | r_1,\omega)] \cdot n dS \tag{6}$$

We have used the property that $\kappa$ is symmetric, which is satisfied for common materials thanks to the Onsager reciprocity relation [36]. It is easy to see that the integrand is zero for all kinds of boundary conditions. The right-hand side of Eq. (5) is simplified using the sifting property of the Dirac delta function and gives

$$\text{r.h.s} = Q_0[G(r_1 | r_0,\omega) - G(r_0 | r_1,\omega)] = 0 \tag{7}$$

The local reciprocity of Eq. (2), under the four conditions listed below it, is thus proved.

## III. Global reciprocity of heat diffusion through two ports

The above definition of reciprocity does not directly apply to transportation through a system with several ports. For wave propagation, the reciprocity of such processes is often defined as the symmetry of the scattering matrix [39]. Similarly, we can study the reciprocity in terms of the relation between field amplitudes at different ports of a diffusive system.

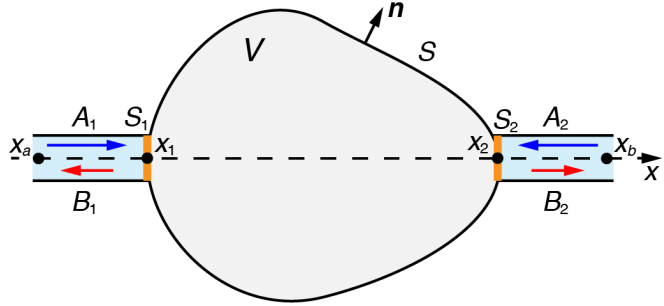

FIG. 2. Global reciprocity of heat transfer through a two-port system

Consider a two-port heat transfer system as in Fig. 2, the region $V$ is thermally insulated, except for two channels $C_1$ and $C_2$ that have uniform and linear material properties. It is also assumed that both channels are long, narrow, and aligned in the lateral $x$ direction, with their upper and lower boundaries insulated. At oscillatory frequency $\omega$, the supported modes are

$$T_c(x,\omega) = e^{\pm ik_c x}, k_c = (1-i)\sqrt{\omega/2D_c} \tag{8}$$

where $D_c$ is the diffusivity of the channels. If the interfaces between the left and right ports and the system are at $x_1$ and $x_2$, we can define the fields in the channels as

$$T_c(x,\omega) = \begin{cases} A_1 e^{-ik_c(x-x_1)} + B_1 e^{ik_c(x-x_1)}, x \leq x_1 \\ A_2 e^{ik_c(x-x_2)} + B_2 e^{-ik_c(x-x_2)}, x \geq x_2 \end{cases} \tag{9}$$

Note that the field evolution is obtained by multiplying $T(x,\omega)$ with $e^{i\omega t}$, so $A_1$ and $A_2$ are the amplitudes of the incident fields, while $B_1$ and $B_2$ are the amplitudes of the outgoing fields. They are related through the scattering matrix $S$

$$\begin{pmatrix} B_1 \\ B_2 \end{pmatrix} = S \begin{pmatrix} A_1 \\ A_2 \end{pmatrix} = \begin{pmatrix} r_{11} & t_{12} \\ t_{21} & r_{22} \end{pmatrix} \begin{pmatrix} A_1 \\ A_2 \end{pmatrix} \tag{10}$$

where $r_{11}$ and $r_{22}$ ($t_{12}$ and $t_{21}$) are the reflection (transmission) coefficients. Reciprocity of scattering processes through the system is defined as a symmetric $S$, or $t_{12} = t_{21}$.

Consider the Green's function for the entire system plus the two channels ($V \cup C_1 \cup C_2$). It satisfies the Neumann boundary condition except for the two ends at $x = \pm L$, and must have the similar form as Eq. (9) in the two channels

$$G(x|x_a,\omega) = \begin{cases} A_1(x_a,\omega)\dfrac{e^{-ik_c(x-x_1)}-e^{ik_c x_1}}{1-e^{ik_c x_1}} + B_1(x_a,\omega)\dfrac{e^{ik_c(x-x_1)}-e^{-ik_c x_1}}{1-e^{-ik_c x_1}}, & x_a \leq x \leq x_1 \\ B_2(x_a,\omega)\dfrac{e^{-ik_c(x-x_2)}-e^{-ik_c(L-x_2)}}{1-e^{-ik_c(L-x_2)}}, & x \geq x_2 \end{cases}$$

$$G(x|x_b,\omega) = \begin{cases} B_1(x_b,\omega)\dfrac{e^{ik_c(x-x_1)}-e^{-ik_c(L+x_1)}}{1-e^{-ik_c(L+x_1)}}, & x \leq x_1 \\ A_2(x_b,\omega)\dfrac{e^{ik_c(x-x_2)}-e^{-ik_c x_2}}{1-e^{-ik_c x_2}} + B_2(x_b,\omega)\dfrac{e^{-ik_c(x-x_2)}-e^{ik_c x_2}}{1-e^{ik_c x_2}}, & x_2 \leq x \leq x_b \end{cases}$$ (11)

where $x_a \leq x_1$ and $x_b \geq x_2$. We adopt this special form in order to ensure a meaningful linear distribution in the limit of $\omega(k_c) \to 0$, instead of the zero fields for Eq. (9). The boundary condition is $G(\pm L|x_a,\omega) = G(\pm L|x_b,\omega) = 0$ for $L \to \infty$. Therefore

$$t_{12} = \frac{B_1(x_b,\omega)}{A_2(x_b,\omega)} = \frac{G(x_1|x_b,\omega)}{A_2(x_b,\omega)}, \quad t_{21} = \frac{B_2(x_a,\omega)}{A_1(x_a,\omega)} = \frac{G(x_2|x_a,\omega)}{A_1(x_a,\omega)}$$ (12)

When the sources are close to the interfaces: $x_a \to x_1$ and $x_b \to x_2$, Eq. (12) is

$$t_{12} = \frac{G(x_1|x_2,\omega)}{A_2(x_2,\omega)}, \quad t_{21} = \frac{G(x_2|x_1,\omega)}{A_1(x_1,\omega)}$$ (13)

Clearly, the amplitude of an incident field only depends on the source, since the reflections at $x = \pm L \to \pm\infty$ are negligible. It is thus reasonable to assume that $A_1(x_1,\omega) = A_2(x_2,\omega) = A(\omega)$, and the reciprocity of the two-port system is equivalent to

$$G(x_1|x_2,\omega) = G(x_2|x_1,\omega)$$ (14)

It means that the condition of global reciprocity is almost identical as that of local reciprocity, except that the positions of source and target should be on the interfaces between the system and the channels. We note that in the limit of $\omega(k_c) \to 0$, one cannot distinguish the incident, reflected, and transmitted parts of a field, since there is no directionality of field propagation. However, the definition of reciprocity in Eq. (14) is still meaningful as an extension from the cases with nonzero frequency.

### IV. Steady-state global reciprocity and thermal diode

For wave propagation, a non-reciprocal two-port system is almost the same thing as a diode, both refer to an asymmetric transmission. It is worth noting that for a diode, the incident and outgoing waves under consideration are generally required to be at equal frequency, while it is correct to state that the transmission between two modes at different frequencies is non-reciprocal [44]. Here, we focus on the equal-frequency transmission.

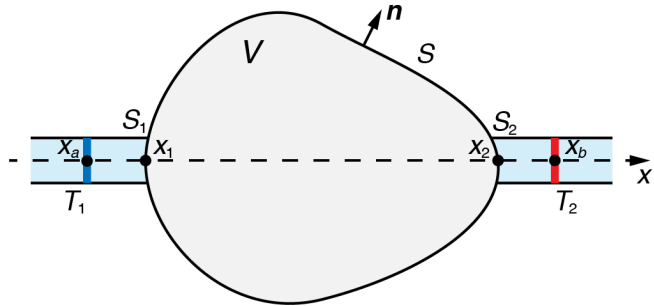

FIG. 3. Thermal diode

In diffusive processes, a diode is usually defined differently. For example, a thermal diode [31]-[33] refers to a system in contact with two thermal reservoirs at temperatures $T_1$ and $T_2$, and the heat fluxes passing through it must be asymmetric after swapping the values of $T_1$ and $T_2$, as shown in Fig. 3. Naturally, one would ask whether a non-reciprocal heat transfer system is still equivalent to a thermal diode. We could answer the question by fixing the temperatures of the interfaces $S_1$ and $S_2$ at $T_1$ ($T_2$).

For such setups, only the fields at zero frequency or steady state are of interest, so we consider the steady-state Green's function. It must be linear in the channels, thus

$$G(x|x_a,0) = \begin{cases} C_1(x_a)(x+L), & -L \leq x \leq x_a \\ D_1(x_a)(x-x_a) + C_1(x_a)(x_a+L), & x_a \leq x \leq x_1 \\ C_2(x_a)(x-L), & x_2 \leq x \leq L \end{cases}$$

$$G(x|x_b,0) = \begin{cases} C_1(x_b)(x+L), & -L \leq x \leq x_1 \\ D_2(x_b)(x-x_b) + C_2(x_b)(x_b-L), & x_2 \leq x \leq x_b \\ C_2(x_b)(x-L), & x_b \leq x \leq L \end{cases} \tag{15}$$

where $x = \pm L$ are the positions of the left and right ends of the channels. Again $x_a \leq x_1$ and $x_b \geq x_2$. The function satisfies Dirichlet boundary condition $G(\pm L|x_a,0) = G(\pm L|x_b,0) = 0$. Assuming that the total heat generation $H = \int h(r)dV$ in $V$ is independent of the conditions outside, the energy conservation requires that

$$\begin{aligned} -\kappa_c D_1(x_a)\sigma &= -\kappa_c C_2(x_a)\sigma + H \\ -\kappa_c C_1(x_b)\sigma &= -\kappa_c D_2(x_b)\sigma + H \\ -\kappa_c [C_1(x_a) - C_2(x_a)]\sigma &= -\kappa_c [C_1(x_b) - C_2(x_b)]\sigma = Q_0 + H \end{aligned} \tag{16}$$

where $\kappa_c$ is the thermal conductivity, and $\sigma$ is the cross-section area of the channels. If the differential operator on the field in region $V$ is linear, the linear combination $F(x) = f_a G(x|x_a,0) + f_b G(x|x_b,0)$ is also a solution to the equation in region $V$. If we require $F(x_a) = T_1$ and $F(x_b) = T_2$, the coefficients satisfy

$$\boldsymbol{G}\begin{pmatrix} f_a \\ f_b \end{pmatrix} = \begin{pmatrix} G(x_a|x_a,0) & G(x_a|x_b,0) \\ G(x_b|x_a,0) & G(x_b|x_b,0) \end{pmatrix}\begin{pmatrix} f_a \\ f_b \end{pmatrix} = \begin{pmatrix} T_1 \\ T_2 \end{pmatrix} \tag{17}$$

In the limit $L \to \infty$, the third equation of Eq. (16) gives

$$\boldsymbol{G}_{11} + \boldsymbol{G}_{12} = \boldsymbol{G}_{22} + \boldsymbol{G}_{21} \tag{18}$$

The heat fluxes $q_1$ and $q_2$ in $C_1$ and $C_2$ near the interfaces $S_1$ and $S_2$ are

$$\begin{pmatrix} q_1 \\ q_2 \end{pmatrix} = \kappa_c \begin{pmatrix} D_1(x_a) & C_1(x_b) \\ C_2(x_a) & D_2(x_b) \end{pmatrix}\begin{pmatrix} f_a \\ f_b \end{pmatrix} = \boldsymbol{Q}\begin{pmatrix} f_a \\ f_b \end{pmatrix} \tag{19}$$

In the limit $L \to \infty$, we have

$$\begin{aligned} \boldsymbol{Q}_{12} &= \kappa_c \boldsymbol{G}_{21}/L \\ \boldsymbol{Q}_{21} &= -\kappa_c \boldsymbol{G}_{12}/L \end{aligned} \tag{20}$$

the first two equations of Eq. (16) give

$$\boldsymbol{Q}_{11} + \boldsymbol{Q}_{22} = \boldsymbol{Q}_{12} + \boldsymbol{Q}_{21} \tag{21}$$

Now we can swap the values of $T_1$ and $T_2$ to solve another set of coefficients

$$\boldsymbol{G}\begin{pmatrix} \tilde{f}_a \\ \tilde{f}_b \end{pmatrix} = \begin{pmatrix} T_2 \\ T_1 \end{pmatrix} = \begin{pmatrix} 0 & 1 \\ 1 & 0 \end{pmatrix}\begin{pmatrix} T_1 \\ T_2 \end{pmatrix} \tag{22}$$

Therefore,

$$\begin{pmatrix} \tilde{f}_a \\ \tilde{f}_b \end{pmatrix} = \boldsymbol{G}^{-1}\begin{pmatrix} 0 & 1 \\ 1 & 0 \end{pmatrix}\boldsymbol{G}\begin{pmatrix} f_a \\ f_b \end{pmatrix} \tag{23}$$

The corresponding heat fluxes are

$$\begin{pmatrix} \tilde{q}_1 \\ \tilde{q}_2 \end{pmatrix} = \bm{Q} \begin{pmatrix} \tilde{f}_a \\ \tilde{f}_b \end{pmatrix} = \bm{Q}\bm{G}^{-1} \begin{pmatrix} 0 & 1 \\ 1 & 0 \end{pmatrix} \bm{G}\bm{Q}^{-1} \begin{pmatrix} q_1 \\ q_2 \end{pmatrix} \tag{24}$$

Note that when $H = 0$, $\bm{Q}$ is singular, but the following results can be similarly obtained.

If the system is not a diode, it is required that

$$\begin{pmatrix} \tilde{q}_1 \\ \tilde{q}_2 \end{pmatrix} = -\begin{pmatrix} q_2 \\ q_1 \end{pmatrix} = -\begin{pmatrix} 0 & 1 \\ 1 & 0 \end{pmatrix} \begin{pmatrix} q_1 \\ q_2 \end{pmatrix} \tag{25}$$

Substituting into Eq. (24) and considering that the temperature values are arbitrarily chosen, we have

$$\bm{Q}\bm{G}^{-1}\begin{pmatrix} 0 & 1 \\ 1 & 0 \end{pmatrix}\bm{G}\bm{Q}^{-1} + \begin{pmatrix} 0 & 1 \\ 1 & 0 \end{pmatrix} = 0 \tag{26}$$

Assuming that the system is reciprocal, namely $\bm{G}_{12} = \bm{G}_{21}$. In the limit $L \to \infty$, Eq. (18) gives that $\bm{G}_{11} = \bm{G}_{22}$. From Eq. (20) and (21) we see that $\bm{Q}$ is anti-symmetric and traceless. Using all the properties, we found that Eq. (26) is satisfied and the system is not a diode. On the other hand, if Eq. (26) is satisfied, we have

$$\begin{aligned}(\bm{G}_{12} - \bm{G}_{21})(\bm{G}_{12} + \bm{G}_{21} - \bm{G}_{11} + \bm{Q}_{11}L/\kappa_c) = 0 \\ (\bm{G}_{12} - \bm{G}_{21})(\bm{G}_{12} - \bm{G}_{21} + \bm{G}_{11} + \bm{Q}_{11}L/\kappa_c) = 0\end{aligned} \tag{27}$$

The only physical solution is $\bm{G}_{12} = \bm{G}_{21}$. We thus prove the equivalence between steady-state global reciprocity and the absence of diode effect, or steady-state global non-reciprocity and being a diode. It is worth mentioning that with heat generation inside the system, there are multiple choices to define a thermal diode, except for the violation of Eq. (25). For example, it can be defined based on the average heat flux $(q_1 + q_2)/2$. However, as we shown above, Eq. (25) is direct related with the global reciprocity, thereby reflecting the essential properties of the system.

## V. Potential mechanisms

*Nonlinearity*

Based on our analysis, it is noticed that there are several methods to break the diffusive reciprocity or make a thermal diode. First, if the governing differential equation is nonlinear, *e.g.* because of a temperature-dependent thermal conductivity or structure [29], its solution cannot be obtained through the Green's function. In such cases, the local and global reciprocity are not very meaningful, while diode effect can be expected.

To demonstrate the effect, consider nonlinear materials 1 and 2 with thermal conductivities $\kappa_1(T) = 0.1 + (T - 275)^3/500$ (W/(m·K)) and $\kappa_2(T) = 0.1 + (325 - T)^3/500$ (W/(m·K)), as plotted in Fig. 4(a). We first just consider material 1, which is used to build a geometrically asymmetric system as shown in Fig. 4(b). On its left and right sides, two channels of size 2 cm × 1 cm are attached. The thermal conductivity of the channels is set as 400 W/(m·K) (that of copper). The left (right) boundary of the left (channel) is maintained at constant temperature $T_1$ ($T_2$).

We perform finite-element simulations using COMSOL Multiphysics on the model with the average temperature $T_0 = (T_1 + T_2)/2 = 300$ K unchanged, and the temperature difference $\Delta T = T_1 - T_0 = T_0 - T_2$ from −25 K to 25 K. Therefore, when $\Delta T > 0$ ($< 0$), heat is transferred in the forward (backward) direction. Also, the amount of heat transferred $Q(\Delta T)$ is positive (negative) in the forward (backward) direction. Since more parts of the system can be heated from the right side, the amount of heat in forward direction should be smaller than that in backward direction, as confirmed by the simulations. However, the rectification ratio is very small: $|Q(-\Delta T)|/|Q(\Delta T)| = 100.48\%$ when $\Delta T = 25$ K. To improve the performance, the configuration in Fig. 4(c) is often used [31]. It is the combination of material 1 and 2 with opposite temperature dependence. Therefore, both parts will have large (small) thermal conductivities for forward (backward) heat transfer. Fig. 4(d) shows the amount of heat transferred through the three systems. The asymmetric nonlinearity makes a thermal diode with more significant rectification ratio: $|Q(\Delta T)|/|Q(-\Delta T)| = 184.32\%$ when $\Delta T = 25$ K. Combining nonlinearity with asymmetry is a well-known strategy for building a thermal diode [29]. It has been shown that simply using nonlinearity is not enough [48],[49]. For one-dimensional heat diffusion, it is also required that the spatial- and temperature-dependence of the thermal conductivity should be nonseparable [50]. Detailed analytical studies on the rectification ratio using temperature-dependent thermal conductivities have been performed recently [51],[52].

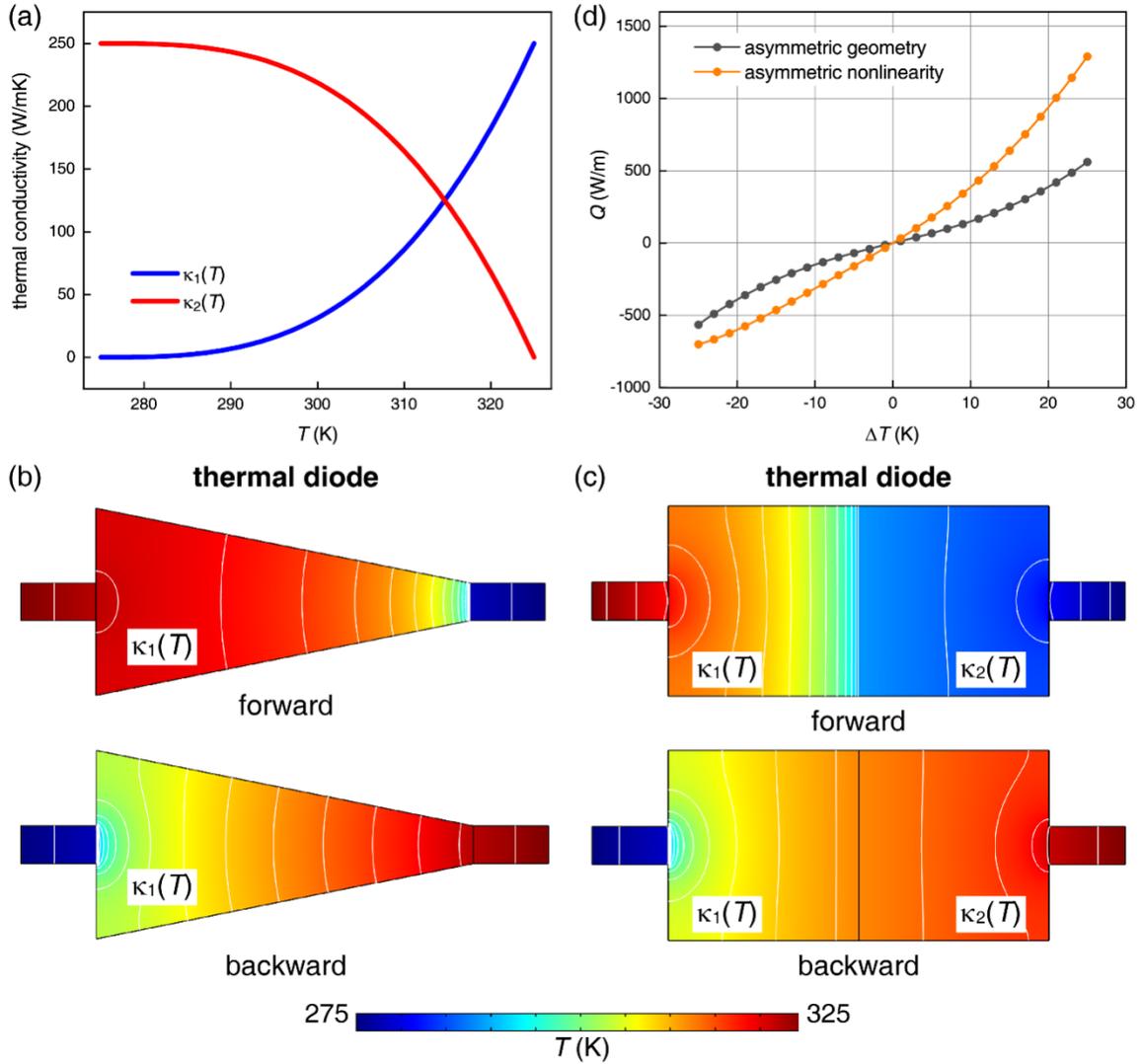

FIG. 4. Effects of nonlinearity. (a) Nonlinear materials 1 and 2 with temperature dependent thermal conductivities $\kappa_1(T)$ and $\kappa_2(T)$. (b) Steady-state temperature profiles for a system built of material 1 with asymmetric geometry. The forward ($\Delta T = 25$ K) and backward ($\Delta T = -25$ K) cases give slightly different amounts of heat. (c) Steady-state temperature profiles for a system built of materials 1 and 2. The forward ($\Delta T = 25$ K) and backward ($\Delta T = -25$ K) cases give different amounts of heat. (d) Heat (per unit thickness) across the system.

*Asymmetric thermal conductivity tensor*

Second, when the thermal conductivity tensor is asymmetric, such as the thermal Hall effect in magnetic fields [53], an additional term appears in Eq. (6)

$$\int_V \nabla G(r|r_1,\omega) \cdot (\kappa - \kappa^T) \cdot \nabla G(r|r_0,\omega) dV \tag{28}$$

which could, but not necessarily be nonzero. To see this, we rewrite Eq. (28) using the property of $\kappa - \kappa^T$, which is an anti-symmetric tensor in three-dimension. In Cartesian coordinate system, it can be expressed as

$$[\kappa - \kappa^T] = \begin{bmatrix} 0 & -a_3 & a_2 \\ a_3 & 0 & -a_1 \\ -a_2 & a_1 & 0 \end{bmatrix} \tag{29}$$

Therefore, we can define a vector $a$ such that its three Cartesian components are $a_1$, $a_2$, and $a_3$. Eq. (28) becomes

$$\begin{aligned}&\int_V \nabla G(r|r_1,\omega) \cdot a \times \nabla G(r|r_0,\omega) dV \\ &= \int_V \nabla \cdot [G(r|r_1,\omega) a \times \nabla G(r|r_0,\omega)] dV - \int_V G(r|r_1,\omega) \nabla \cdot [a \times \nabla G(r|r_0,\omega)] dV\end{aligned} \tag{30}$$

If only Dirichlet boundary condition is applied, the first term is zero based on Gauss's law. In the second term, we have

$$\nabla \cdot [a \times \nabla G(r|r_0,\omega)] = \nabla G(r|r_0,\omega) \cdot (\nabla \times a) - a \cdot [\nabla \times \nabla G(r|r_0,\omega)] = \nabla G(r|r_0,\omega) \cdot (\nabla \times a) \tag{31}$$

As a result, the additional term is still zero for uniform materials, which obviously also applies for two-dimensional cases. When Neumann or mixed boundary condition exists, the first term in Eq. (30) does not vanish. We thus shown that the local reciprocity in a uniform material with asymmetric thermal conductivity tensor is preserved when all the boundaries are at fixed temperatures.

To confirm the result, we perform finite-element simulations on a rectangular region (size 10 cm × 5 cm, thickness 1 mm) built of uniform material with thermal conductivity $\kappa$. In the Cartesian coordinate system whose origin is at the center of the region, the matrix form of $\kappa$ is

$$[\kappa] = \kappa_0 \begin{bmatrix} 1 & 0.3 \\ -0.3 & 1 \end{bmatrix} \tag{32}$$

where $\kappa_0 = 100$ W/(m·K). A heat source $Q = 5$ W is uniformly launched on a small spot of radius 0.3 mm at $[r_0] = [-2 \text{ cm } 0.5 \text{ cm}]^T$ to mimic a point heat source. If all boundaries of the region are maintained at constant temperature 300 K as in Fig. 5(a), the temperature at $[r_1] = [1 \text{ cm } -1 \text{ cm}]^T$ is 301.7 K. After swapping the positions of $r_0$ and $r_1$, this temperature value remains the same. On the other hand, if only the left side is maintained at constant temperature with the other sides thermally insulated as in Fig. 5(b), the temperature at $r_1$ is 326.90 K. After swapping the positions of $r_0$ and $r_1$, this temperature becomes 329.38 K, indicating a broken local reciprocity due to the existence of Neumann boundary conditions. For comparison, a symmetric thermal conductivity $\kappa = \kappa_0$ gives the same temperature 329.60 K before and after swapping the positions of $r_0$ and $r_1$.

We also simulated the cases where all boundaries are thermally insulated as in Fig. 5(c). In such a system steady-state solution does not exist for a constant point heat source, so we consider the time-harmonic solutions with $Q = 5\cos(\omega t)$ (W), where $\omega = 2\pi \times 0.001$ rad/s. The density and heat capacity of the material are set as $\rho = 8000$ kg/m³ and $c_p = 500$ J/(kg·K). It turns out that the temperature variation amplitude at $r_1$ is $A_1 = 39.03$ K, while the value becomes $A_2 = 39.27$ K after swapping the positions of $r_0$ and $r_1$. This again confirms that the local reciprocity is broken for Neumann boundary conditions. For comparison, a symmetric thermal conductivity $\kappa = \kappa_0$ gives the same temperature amplitude 39.09 K before and after swapping the positions of $r_0$ and $r_1$. Of course, when the thermal conductivity is nonuniform and $\nabla \times \boldsymbol{a}$ is nonzero, the local nonreciprocity is broken following Eq. (31), irrespective of the boundary conditions.

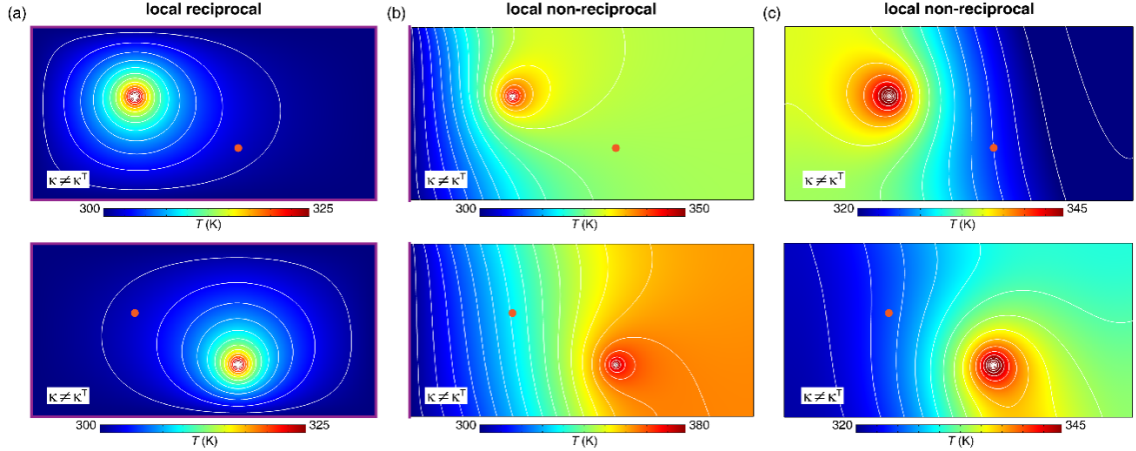

FIG. 5. Effects of asymmetric thermal conductivity tensor: local reciprocity. (a) Steady-state temperature profiles generated by a point heat source with all boundaries at fixed temperature. The reciprocity is preserved. (b) Steady-state temperature profiles generated by a point heat source with the left side at fixed temperature and the other three boundaries thermally insulated. The reciprocity is broken. (c) Temperature profiles at $t = 9120$ s generated by a time-harmonic point heat source with all boundaries thermally insulated. The reciprocity is broken. Boundaries with fixed temperatures are indicated with purple lines. The target positions where temperatures are measured are indicated by orange dots.

Next, we study the global reciprocity, especially that at steady state to determine whether the system could become a thermal diode. In fact, it can be shown that at zero frequency, simply using asymmetric thermal conductivity tensor does not break the global reciprocity of a two-port system, so a thermal diode cannot be built this way. To prove this, we use the same setup as in Fig. 2. Similar to Eq. (5), an integral on the system gives

$$\int_V \{G(r|r_1,\omega)\nabla \cdot [\boldsymbol{\kappa} \cdot \nabla G(r|r_0,\omega)] - G(r|r_0,\omega)\nabla \cdot [\boldsymbol{\kappa} \cdot \nabla G(r|r_1,\omega)]\} dV$$
$$= \int_V [G(r|r_0,\omega)Q_0\delta(r-r_1) - G(r|r_1,\omega)Q_0\delta(r-r_0)] dV = 0 \tag{33}$$

Now that $r_0$ and $r_1$ are outside the system in the channels, so the integral simply vanishes. We can also integral on the system plus the two channels

$$\int_{V\cup C_1 \cup C_2} \{G(r|r_1,\omega)\nabla \cdot [\boldsymbol{\kappa} \cdot \nabla G(r|r_0,\omega)] - G(r|r_0,\omega)\nabla \cdot [\boldsymbol{\kappa} \cdot \nabla G(r|r_1,\omega)]\} dV$$
$$= \int_{V\cup C_1 \cup C_2} [G(r|r_0,\omega)Q_0\delta(r-r_1) - G(r|r_1,\omega)Q_0\delta(r-r_0)] dV = Q_0[G(r_1|r_0,\omega) - G(r_0|r_1,\omega)] \tag{34}$$

Using Eq. (33), the left-hand side is simplified as

$$\text{l.h.s} = \int_{C_1 \cup C_2} \{G(r|r_1,\omega)\nabla \cdot [\kappa_c \nabla G(r|r_0,\omega)] - G(r|r_0,\omega)\nabla \cdot [\kappa_c \nabla G(r|r_1,\omega)]\} dV$$
$$= \int_{\partial C_1 \cup \partial C_2} \kappa_c [G(r|r_1,\omega)\nabla G(r|r_0,\omega) - G(r|r_0,\omega)\nabla G(r|r_1,\omega)] \cdot \boldsymbol{n} dS \tag{35}$$

On the two interfaces $S_1$ and $S_2$ between the channels and the system. At zero frequency ($\omega = 0$), energy conservation requires that

$$\int_{S_1} \kappa_c \nabla G(\mathbf{r}\,|\,\mathbf{r}_0,0) \cdot \mathbf{n}\, dS + \int_{S_2} \kappa_c \nabla G(\mathbf{r}\,|\,\mathbf{r}_0,0) \cdot \mathbf{n}\, dS = 0$$
$$\int_{S_1} \kappa_c \nabla G(\mathbf{r}\,|\,\mathbf{r}_1,0) \cdot \mathbf{n}\, dS + \int_{S_2} \kappa_c \nabla G(\mathbf{r}\,|\,\mathbf{r}_1,0) \cdot \mathbf{n}\, dS = 0$$
(36)

Combining with the other boundary conditions of $C_1$ and $C_2$, the surface integral in Eq. (35) vanishes. Therefore, the right-hand side of Eq. (34) equals zero. We thus prove that the steady-state global reciprocity is preserved even for a system with asymmetric thermal conductivity tensor, which could even be nonuniform. Based on the equivalence shown in Section IV, it cannot become a thermal diode. Of course, at nonzero frequencies we do not have Eq. (36), so the global reciprocity is broken.

To confirm the above results, we perform finite-element simulations on a system with the same shape as in Fig. 5. Two channels of size 2 cm × 1 cm with thermal conductivity $\kappa_c$ = 400 W/(m·K), density $\rho_c$ = 8900 kg/m$^3$, and heat capacity $c_{pc}$ = 390 J/(kg·K) (those of copper) are attached to the system. Channel 1 is connected to the left of the system. Channel 2 is connected to the bottom of the system. The thermal conductivity of the system follows Eq. (32). The global reciprocity is tested by applying an oscillatory point heat source as for results in Fig. 5(c), with all boundaries thermally insulated. When the point heat source is in channel 1 (5 mm from the interface), the amplitude of temperature variation in channel 2 (5 mm from the interface) is $A_1$ = 38.89 K. The value becomes $A_2$ = 42.06 K after the positions of heat source and target are swapped. The temperature distributions at $t$ = 9110 s are shown in Fig. 6(a). However, the ratio $A_2/A_1$ decreases toward unity as the oscillating frequency $\omega$ approaches zero, as shown in Fig. 6(b). It confirms our theoretical prediction. We also apply fixed boundary conditions on the left end of channel 1 and lower end of channel 2, whose temperatures $T_1 = T_0 + \Delta T$ and $T_2 = T_0 - \Delta T$ are set the same as for the nonlinear cases. As shown in Fig. 6(c), the temperature distributions for the forward and backward cases are symmetric. As expected, the heat is linearly proportional to $\Delta T$, so $Q(\Delta T) = -Q(-\Delta T)$ and the system is not a thermal diode [Fig. 6(d)].

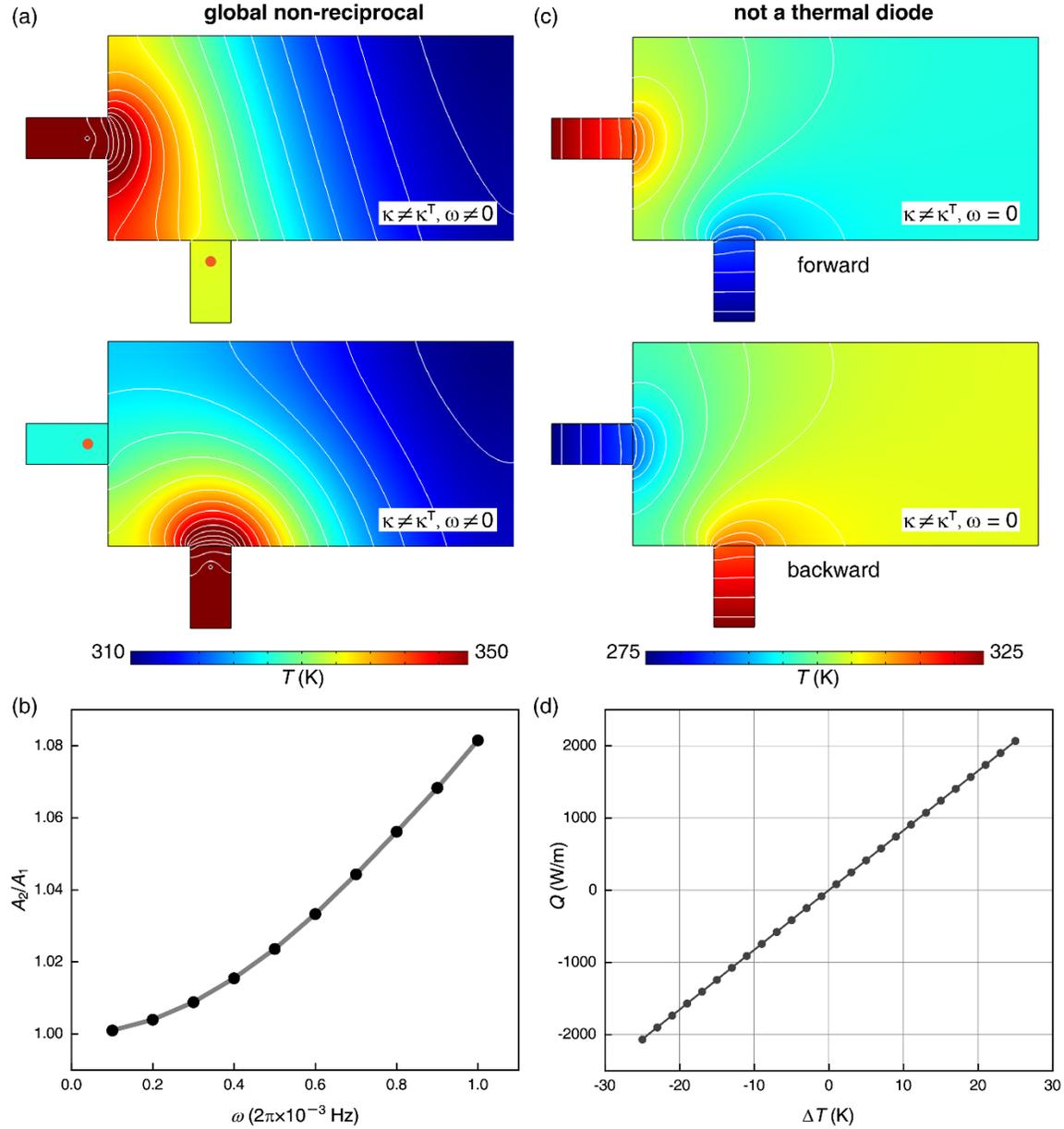

FIG. 6. Effects of asymmetric thermal conductivity tensor: global reciprocity and diode effect. (a) Temperature profiles at $t = 9110$ s generated by a time-harmonic point heat source with all boundaries thermally insulated. The global reciprocity is broken. (b) As the oscillating frequency $\omega$ of the heat source approaches zero, the ratio between the two amplitudes approaches one, indicating a preserved global reciprocity at steady state. (c) Steady-state temperature profiles with the left and lower boundaries at fixed temperatures. The forward ($\Delta T = 25$ K) and backward ($\Delta T = -25$ K) cases are symmetric, so the system is not a thermal diode. (d) Heat (per unit thickness) across the system. The target positions where temperatures are measured are indicated by orange dots.

*Thermal convection*

Third, if thermal convection (assuming forced convection with temperature-independent velocity field) is present, an additional term appears that could possibly break the local reciprocity

$$-\int_{V}\left[G(r\,|\,r_{1},\omega)\rho c_{p}\mathbf{v}\cdot\nabla G(r\,|\,r_{0},\omega)-G(r\,|\,r_{0},\omega)\rho c_{p}\mathbf{v}\cdot\nabla G(r\,|\,r_{1},\omega)\right]dV \qquad (37)$$

Such a convection induced non-reciprocity has been studied for mechanically rotating structures, whose effective thermal conductivities tensor are shown to be asymmetric [54]. The non-reciprocity is reflected in a rotated temperature field around the structure. Interestingly, the reciprocity can be restored by combining counter-rotating parts, thereby disguising a convective system as a purely conductive one [55]. It should be noted that this reciprocity is between any two arbitrary directions, so it is basically a local one. In terms of a two-port setup, the result is quite different. Based on the effective thermal conductivity [54], most results for asymmetric thermal conductivity tensor are also valid for heat convection through rotating structures. Namely, the local reciprocity is broken; the global reciprocity for a two-port system is broken for nonzero frequencies but preserved at zero frequency based on the similar argument for Eq. (34); the system cannot become a thermal diode. Note that Eq. (37) cannot be manipulated like Eq. (28), so the local reciprocity is broken even for pure Dirichlet boundary conditions.

All the predictions are numerically demonstrated for a 1 mm thick solid ring with inner and outer radius 1.5 cm and 3 cm rotating at angular speed $\Omega = 2\pi \times 0.002$ rad/s. Its thermal conductivity, density, and heat capacity are set as $\kappa = 16$ W/(m·K), $\rho = 8000$ kg/m$^3$, and $c_p = 500$ J/(kg·K) (those of stainless steel). For local reciprocity, all the boundaries are thermally insulated. In a polar coordinate system $(r, \theta)$ whose origin is built at the center of the system, a heat source $Q = 2\cos(\omega t)$ (W) is placed at $[r_0] = [2.25$ cm $5\pi/4]^T$. For $\omega = 2\pi \times 0.001$ rad/s, the amplitude of temperature variation at $[r_1] = [2.25$ cm $7\pi/4]^T$ is $A_1 = 52.23$ K. The value becomes $A_2 = 37.46$ K after the positions of $r_0$ and $r_1$ are swapped. The local reciprocity is broken, as shown in Fig. 7(a). The same is confirmed for fixed boundary conditions. The temperature fields are not shown as they are not very informative: only the vicinity of the heat source is heated up.

To test the global reciprocity, two channels are attached to the ring at $\theta = 5\pi/4$ and $7\pi/4$ as in Fig. 7(b). The size and material parameters of the channels are the same as used in above simulations. The heat source $Q$ is placed at $[r_0] = [3.5$ cm $5\pi/4]^T$. The amplitude of temperature variation at $[r_1] = [3.5$ cm $7\pi/4]^T$ is $A_1 = 43.28$ K. The value becomes $A_2 = 31.17$ K after the positions of $r_0$ and $r_1$ are swapped, indicating global non-reciprocity. It is also verified that as $\omega \to 0$, the ratio $A_1/A_2$ approaches unity, which confirms the steady-state global reciprocity. The system is thus not a thermal diode as shown in Fig. 7(c), where outer end of channel 1 (2) is maintained at fixed temperature $T_1 = T_0 + \Delta T$ ($T_2 = T_1 = T_0 - \Delta T$), same as for the nonlinear cases. The temperature profiles for forward and backward cases are obviously symmetric. Since there is no heat generation in the system, the amount of heat through both channels should equal. We thus choose to calculate the heat $Q(\Delta T)$ as an integral over a cross-section in the right channel (yellow lines in Fig. 7(c)). The results for different rotation speeds $\Omega$ of the ring are plotted in Fig. 7(d). All of them are odd symmetric to $\Delta T$, indicating the absence of diode effect.

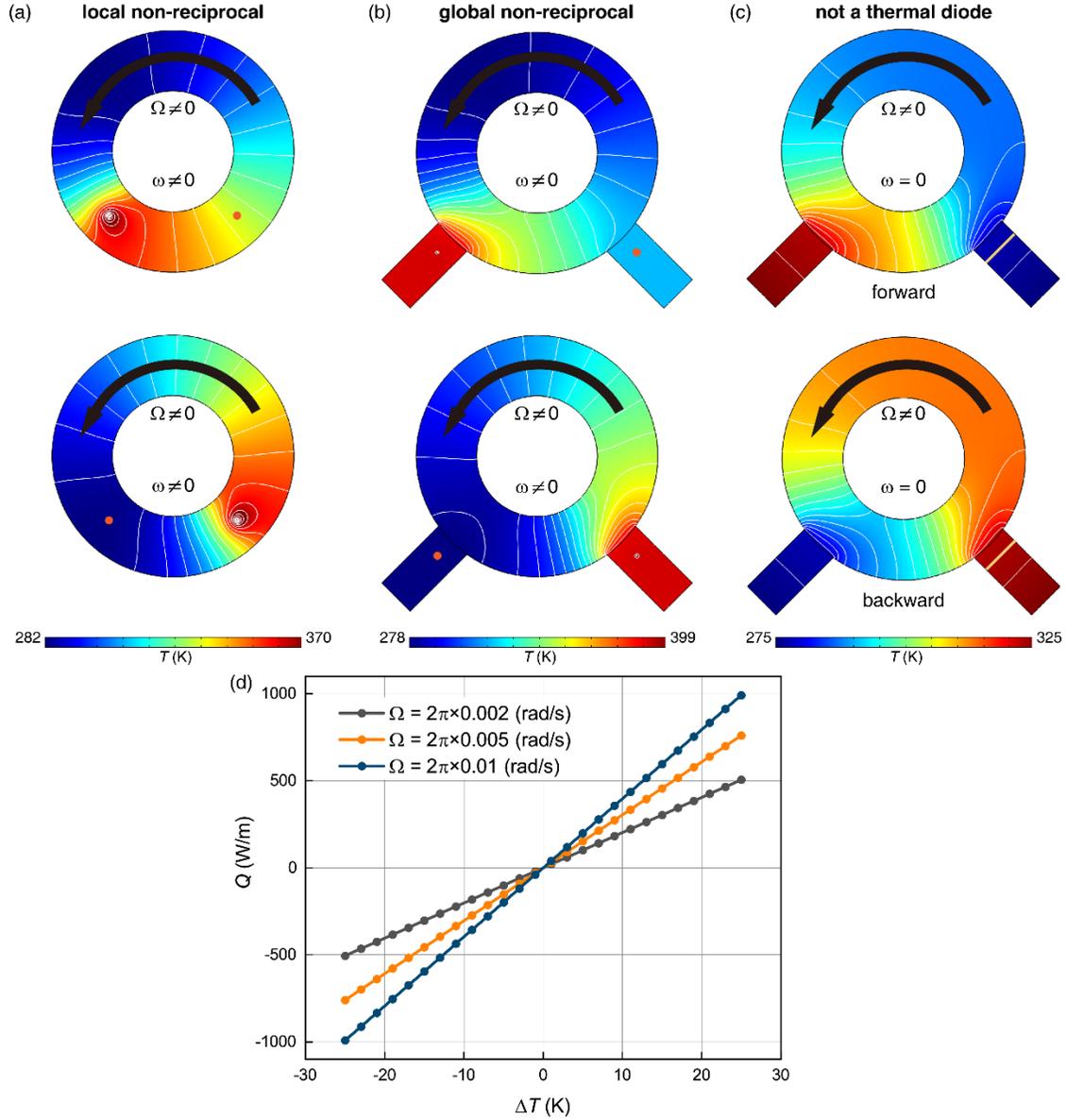

FIG. 7. Effects of heat convection. (a) Temperature profiles on a rotating ring (with rotation speed Ω) at $t$ = 9110 s generated by a time-harmonic point heat source (at angular frequency $\omega$) with all boundaries thermally insulated. The local reciprocity is broken. (b) Temperature profiles on a rotating ring attached with two channels at 9080 s generated by a time-harmonic point heat source with all boundaries thermally insulated. The global reciprocity is broken. (c) Steady-state temperature profiles with the left and lower boundaries at fixed temperatures. The forward ($\Delta T$ = 25 K) and backward ($\Delta T$ = −25 K) cases are symmetric, so the system is not a thermal diode. The target positions where temperatures are measured are indicated by orange dots. (d) Heat (per unit thickness) across the system at different rotation speeds Ω.

*Heat generation*

In our analysis, there could be heat generation inside the system, so its scope is broader than in many other works on thermal diode and rectification. As proved in Section IV, when the total heat generation $H$ is independent of the outside thermal conditions, the equivalence between steady-state global reciprocity and diode effect is still valid. In addition, it is easy to see that when the distribution of heat generation density $h(r)$ in the system is inhomogeneous and asymmetric, it can break the steady-state global reciprocity without relying on other effects. Therefore, inhomogeneous heat generation is a convenient mechanism to make a thermal diode.

Here, we further consider the case when there exists a heat source in the system that depends on the temperatures in the channels. Eq. (16) and the following derivations are incorrect as $H$ is temperature-dependent. Such a case can be easily realized using a setup as shown in Fig. 8(a). The system is a rectangular of size 10 cm × 5 cm (thickness 1mm) with thermal conductivity $\kappa = 100$ W/(m·K). A square hole of size 1 cm × 1 cm is made inside. The center of the square is 1 cm left from the center of the rectangular. Two channels of size 2 cm × 1 cm with thermal conductivity $\kappa_c = 400$ W/(m·K) are attached to the left and right boundaries of the system. To introduce a temperature-dependent heat source, the boundaries of the square hole are maintained at fixed temperature $T_0 = 300$ K. Since this is simply Dirichlet boundary condition, the local and global reciprocities are preserved following the proofs in Section II and III. The global reciprocity is numerically confirmed [Fig. 8(a)] by applying a heat source $Q = 5$ W to a small spot of radius 0.3 mm at the center of channel 1. The temperature at the center of channel 2 is 303.33 K. The value remains the same after swapping the positions of the heat source and target.

Despite of the steady-state global reciprocity the system violates Eq. (25) [Fig. 8(b)], because the heat generation inside depends on the temperature difference between $T_0$ and $T_1$ ($T_2$). As the hole is closer to channel 1, the difference between $T_0$ and $T_1$ contributes more. Therefore, in the forward (backward) case, the heat generation is negative (positive) as $T_0$ is lower (higher) than $T_1$, namely, $Q_1(\Delta T) - Q_2(\Delta T) > 0$ for $\Delta T > 0$ and $Q_1(\Delta T) - Q_2(\Delta T) < 0$ for $\Delta T < 0$, where $Q_1$ and $Q_2$ are the heat in channel 1 and 2, respectively. The linearity and symmetric boundary conditions also imply that $Q_1(-\Delta T) = -Q_1(\Delta T)$ and $Q_2(-\Delta T) = -Q_2(\Delta T)$. It follows that $Q_1(\Delta T) + Q_2(-\Delta T) \neq 0$ or the average heat $[Q_1(\Delta T) + Q_2(\Delta T)]/2 \neq -[Q_1(-\Delta T) + Q_2(-\Delta T)]/2$ [Fig. 8(c)]. Strictly speaking, the system is a three-port one with the central hole in contact with some thermal reservoir, so it does not meet the rigorous definition of a thermal diode. However, it is still a representative counterexample for the equivalence between Eq. (25) and steady-state global reciprocity.

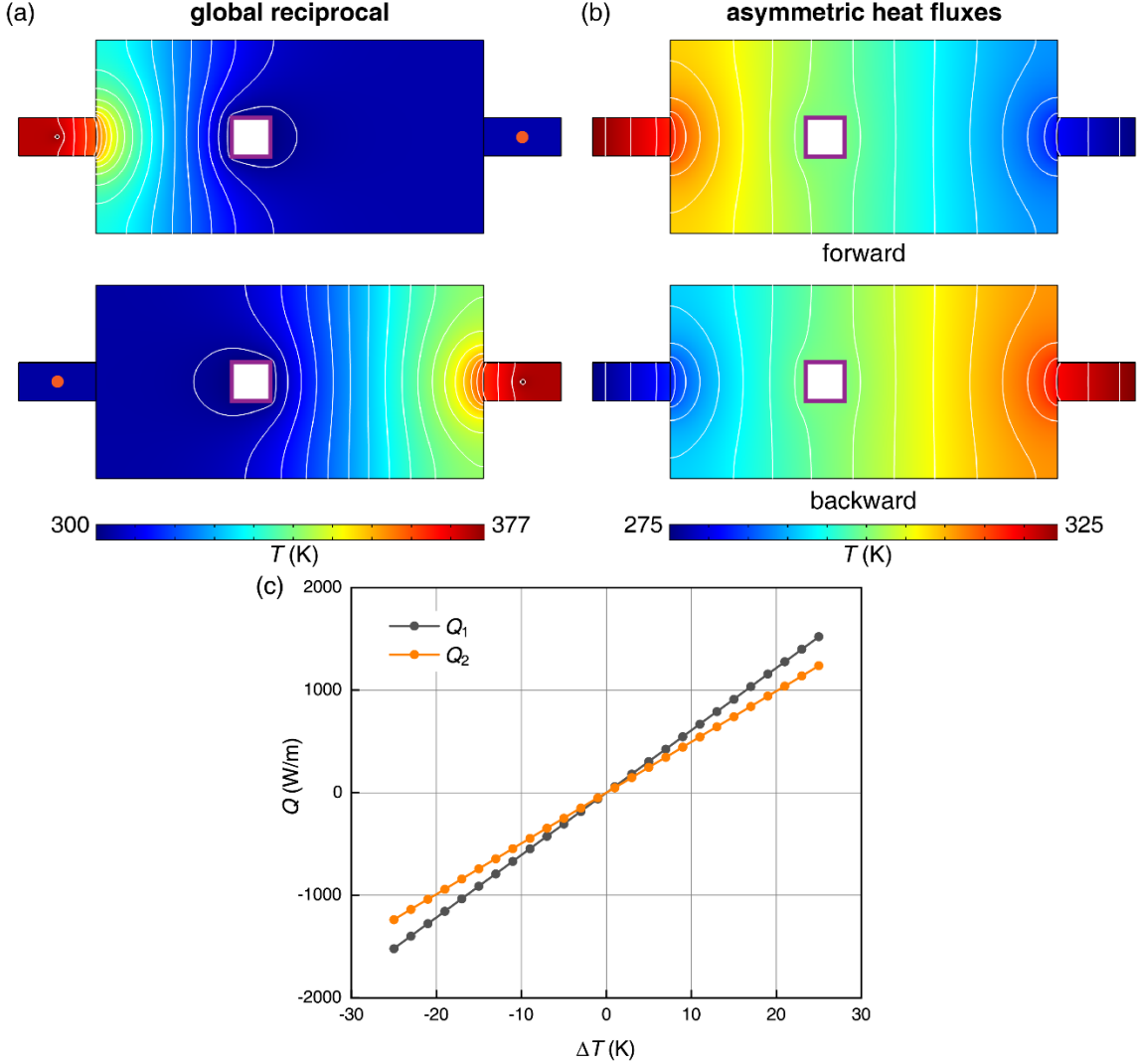

FIG. 8. Effects of temperature-dependent heat source. (a) Temperature profiles on a two-port system with an interior hole at fixed temperature generated by a point heat source with all boundaries thermally insulated. The global reciprocity is preserved. (b) Steady-state temperature profiles with the left and right boundaries at fixed temperatures. The forward ($\Delta T = 25$ K) and backward ($\Delta T = -25$ K) cases are asymmetric. (c) Heat (per unit thickness) across the system. Boundaries with fixed temperatures are indicated with purple lines. The target positions where temperatures are measured are indicated by orange dots.

## VI. Conclusion

Based on the frequency-domain Green's function, we have provided a reasonable definition of scattering matrix and global reciprocity of diffusion which extends to the zero-frequency limit. We also proved that for linear systems without temperature-dependent heat generation, steady-state global reciprocity indicates the absence of diode effect, while breaking the global reciprocity makes a thermal diode. We discussed several potential mechanisms that may break the reciprocity and make a thermal diode. Nonlinearity with asymmetric geometry makes a thermal diode. Asymmetric thermal conductivity tensor breaks the local reciprocity and global reciprocity at non-zero frequency. However, for a two-port system, the steady-state global reciprocity is preserved, so asymmetric thermal conductivity tensor does not make a thermal diode. Typical systems with heat convection and temperature-dependent heat generation are also studied. They also preserve steady-state global reciprocity, but the forward and backward heat fluxes could be assymetric in the latter. Finally, dynamic materials whose parameters are time-modulated [56] have attracted great interests as a potential tool to induce diffusive non-reciprocity [57],[58]. We will discuss about it in a separate work.